\input harvmac
\input amssym

\def\det{{\rm det}}


\def\IL{\relax{\rm I\kern-.18em L}}
\def\IH{\relax{\rm I\kern-.18em H}}
\def\IR{\relax{\rm I\kern-.18em R}}
\def\IC{\relax\hbox{$\inbar\kern-.3em{\rm C}$}}
\def\IZ{\relax\ifmmode\mathchoice
{\hbox{\cmss Z\kern-.4em Z}}{\hbox{\cmss Z\kern-.4em Z}}
{\lower.9pt\hbox{\cmsss Z\kern-.4em Z}} {\lower1.2pt\hbox{\cmsss
Z\kern-.4em Z}}\else{\cmss Z\kern-.4em Z}\fi}

\def\CN {{\cal N}}

\def\CO {{\cal O}}


\def\CN {{\cal N}}

\def\CO {{\cal O}}

\def\det{{\rm det}}
\def\Tr{{\rm Tr}}

\font\manual=manfnt \def\dbend{\lower3.5pt\hbox{\manual\char127}}

\def\IZ{\relax\ifmmode\mathchoice
{\hbox{\cmss Z\kern-.4em Z}}{\hbox{\cmss Z\kern-.4em Z}}
{\lower.9pt\hbox{\cmsss Z\kern-.4em Z}} {\lower1.2pt\hbox{\cmsss
Z\kern-.4em Z}}\else{\cmss Z\kern-.4em Z}\fi}
\def\half {{1\over 2}}

\def\lfm#1{\medskip\noindent\item{#1}}

\def\bar{\overline}

\def\rt2{\sqrt{2}}
\def\irt2{{1\over\sqrt{2}}}

\def\slashchar#1{\setbox0=\hbox{$#1$}           
   \dimen0=\wd0                                 
   \setbox1=\hbox{/} \dimen1=\wd1               
   \ifdim\dimen0>\dimen1                        
      \rlap{\hbox to \dimen0{\hfil/\hfil}}      
      #1                                        
   \else                                        
      \rlap{\hbox to \dimen1{\hfil$#1$\hfil}}   
      /                                         
   \fi}

\lref\OSV{ H.~Ooguri, A.~Strominger and C.~Vafa,
  ``Black hole attractors and the topological string,''
  Phys.\ Rev.\ D {\bf 70}, 106007 (2004)
  [arXiv:hep-th/0405146]. }

\lref\VafaQA{
  C.~Vafa,
  ``Two dimensional Yang-Mills, black holes and topological strings,''
  arXiv:hep-th/0406058.
}

\lref\dexact{ A.~Dabholkar,
  ``Exact counting of black hole microstates,''
  arXiv:hep-th/0409148. }

\lref\AganagicJS{
  M.~Aganagic, H.~Ooguri, N.~Saulina and C.~Vafa,
  ``Black holes, q-deformed 2d Yang-Mills, and non-perturbative topological
  strings,''
  Nucl.\ Phys.\ B {\bf 715}, 304 (2005)
  [arXiv:hep-th/0411280].
}

\lref\verlinde{
  E.~Verlinde,
  ``Attractors and the holomorphic anomaly,''
  arXiv:hep-th/0412139;
}

\lref\CardosoXF{
  G.~L.~Cardoso, B.~de Wit, J.~Kappeli and T.~Mohaupt,
  ``Asymptotic degeneracy of dyonic N = 4 string states and black hole
  entropy,''
  JHEP {\bf 0412}, 075 (2004)
  [arXiv:hep-th/0412287].
}

\lref\SenPU{
  A.~Sen,
  ``Black holes, elementary strings and holomorphic anomaly,''
  arXiv:hep-th/0502126.
}

\lref\ddmp{
  A.~Dabholkar, F.~Denef, G.~W.~Moore and B.~Pioline,
  ``Exact and asymptotic degeneracies of small black holes,''
  arXiv:hep-th/0502157.
}

\lref\SenCH{
  A.~Sen,
  ``Black holes and the spectrum of half-BPS states in N = 4 supersymmetric
  string theory,''
  arXiv:hep-th/0504005.
}

\lref\AganagicDH{
  M.~Aganagic, A.~Neitzke and C.~Vafa,
  ``BPS microstates and the open topological string wave function,''
  arXiv:hep-th/0504054.
}

\lref\DijkgraafBP{
  R.~Dijkgraaf, R.~Gopakumar, H.~Ooguri and C.~Vafa,
  ``Baby universes in string theory,''
  arXiv:hep-th/0504221.
}

\lref\SenKJ{
  A.~Sen,
  ``Stretching the horizon of a higher dimensional small black hole,''
  arXiv:hep-th/0505122.
}

\lref\SenWA{
  A.~Sen,
  ``Black hole entropy function and the attractor mechanism in higher
  derivative gravity,''
  arXiv:hep-th/0506177.
}

\lref\PiolineVI{
  B.~Pioline,
  ``BPS black hole degeneracies and minimal automorphic representations,''
  arXiv:hep-th/0506228.
}

\lref\DabholkarDT{
  A.~Dabholkar, F.~Denef, G.~W.~Moore and B.~Pioline,
  ``Precision counting of small black holes,''
  arXiv:hep-th/0507014.
}

\lref\msw{
  J.~M.~Maldacena, A.~Strominger and E.~Witten,
  ``Black hole entropy in M-theory,''
  JHEP {\bf 9712}, 002 (1997)
  [arXiv:hep-th/9711053].
}

\lref\corr{ G.~Lopes Cardoso, B.~de Wit and T.~Mohaupt,
  ``Corrections to macroscopic supersymmetric black-hole entropy,''
  Phys.\ Lett.\ B {\bf 451}, 309 (1999)
  [arXiv:hep-th/9812082]. }

\lref\DabholkarJT{
  A.~Dabholkar and J.~A.~Harvey,
  ``Nonrenormalization Of The Superstring Tension,''
  Phys.\ Rev.\ Lett.\  {\bf 63}, 478 (1989).
}

\lref\DabholkarYF{
  A.~Dabholkar, G.~W.~Gibbons, J.~A.~Harvey and F.~Ruiz Ruiz,
  ``Superstrings And Solitons,''
  Nucl.\ Phys.\ B {\bf 340}, 33 (1990).
}

\lref\ssy{
  D.~Shih, A.~Strominger and X.~Yin,
  ``Recounting dyons in N = 4 string theory,''
  arXiv:hep-th/0505094.
}

\lref\ssyii{
  D.~Shih, A.~Strominger and X.~Yin,
  ``Counting dyons in N = 8 string theory,''
  arXiv:hep-th/0506151.
}

\lref\DVV{
  R.~Dijkgraaf, E.~Verlinde and H.~Verlinde,
  ``Counting dyons in N = 4 string theory,''
  Nucl.\ Phys.\ B {\bf 484}, 543 (1997)
  [arXiv:hep-th/9607026].
}

\lref\GaiottoGF{
  D.~Gaiotto, A.~Strominger and X.~Yin,
  ``New connections between 4D and 5D black holes,''
  arXiv:hep-th/0503217.
}

\lref\vijay{
  V.~Balasubramanian,
  ``How to count the states of extremal black holes in N = 8 supergravity,''
  arXiv:hep-th/9712215.
}

\lref\KalloshUY{
  R.~Kallosh and B.~Kol,
  ``E(7) Symmetric Area of the Black Hole Horizon,''
  Phys.\ Rev.\ D {\bf 53}, 5344 (1996)
  [arXiv:hep-th/9602014].
}

\lref\FerraraUM{
  S.~Ferrara and R.~Kallosh,
  ``Universality of Supersymmetric Attractors,''
  Phys.\ Rev.\ D {\bf 54}, 1525 (1996)
  [arXiv:hep-th/9603090].
}

\lref\HarveyIR{
  J.~A.~Harvey and G.~W.~Moore,
  ``Fivebrane instantons and R**2 couplings in N = 4 string theory,''
  Phys.\ Rev.\ D {\bf 57}, 2323 (1998)
  [arXiv:hep-th/9610237].
}

\lref\NekrasovJS{
  N.~Nekrasov, H.~Ooguri and C.~Vafa,
  ``S-duality and topological strings,''
  JHEP {\bf 0410}, 009 (2004)
  [arXiv:hep-th/0403167].
}

\lref\CremmerUP{
  E.~Cremmer and B.~Julia,
  ``The SO(8) Supergravity,''
  Nucl.\ Phys.\ B {\bf 159}, 141 (1979).
}


\newbox\tmpbox\setbox\tmpbox\hbox{\abstractfont }
\noblackbox
 \Title{\vbox{\baselineskip12pt\hbox to\wd\tmpbox{\hss
}}\hbox{hep-th/0508174 }} {\vbox{\centerline{Exact Black Hole Degeneracies}\medskip\centerline{and the Topological String}\medskip\centerline{} }}

\centerline{David Shih$^a$ and Xi Yin$^b$ }
\smallskip\centerline{$^a$ Department of Physics, Princeton
University, Princeton, NJ 08544, USA} \centerline{$^b$ Jefferson
Physical Laboratory, Harvard University, Cambridge, MA 02138, USA}

\centerline{} \vskip.4in \centerline{\bf Abstract} { Motivated by
the recent conjecture of Ooguri, Strominger and Vafa, we compute
the semi-canonical partition function of BPS black holes in ${\cal
N}=4$ and ${\cal N}=8$ string theories, to all orders in
perturbation theory. Not only are the black hole partition
functions surprisingly simple; they capture the full topological
string amplitudes, as expected from the OSV conjecture. The
agreement is not perfect, however, as there are differences
between the black hole and topological string partition functions
even at the perturbative level. We propose a minimal modification
of the OSV conjecture, in which these differences are understood
as a nontrivial measure factor for the topological string.

 } \vskip.2in

\Date{August, 2005} \vfill

\listtoc \writetoc

\newsec{Introduction}
A simple relation between the partition function of BPS black
holes and that of topological strings, of the form
\eqn\zzs{Z_{BH}(p,\phi) = \sum_q \Omega_{BH}(p,q) e^{-q\cdot\phi} =
|Z_{top}(p+i\phi)|^2,} was recently proposed by Ooguri, Strominger
and Vafa (OSV) in \OSV. Subsequently, much work has been done to
clarify and test this proposal
\refs{\VafaQA\dexact\AganagicJS\verlinde\CardosoXF\SenPU\ddmp\SenCH
\AganagicDH\DijkgraafBP\SenKJ\SenWA\PiolineVI-\DabholkarDT}.
Clarification is definitely needed, as a quick glance at \zzs\
shows that the original proposal of OSV is schematic at best --
taking it literally immediately leads to problems. For instance,
the LHS of \zzs\ is manifestly invariant under integer shifts of
$\phi$, $\phi\to \phi+2\pi i k$, while the RHS is generally not.
Another point that needs clarification is the definition of black
hole partition function $\Omega_{BH}(p,q)$: in general, there are
multiple ``supersymmetric indices" which have the correct
asymptotics and could potentially be used in \zzs.

Since the precise formulation of the OSV proposal is currently
lacking, it is especially important to have examples of black holes
whose degeneracies are (a) known exactly, and (b) sufficiently
simple that the OSV transform $Z_{BH}(p,\phi)$ can be evaluated
explicitly.

The OSV conjecture was originally formulated for BPS black holes
in ${\cal N}=2$ compactifications of string theories. The
classical entropy (and certain subleading corrections) of such
black holes has been understood microscopically \refs{\msw,\corr}.
However, general formulas for the exact degeneracies of large
$\CN=2$ black holes are not known.\foot{The exact degeneracy of
certain classes of small $\CN=2$ black holes is known (see e.g.\
\DabholkarDT). By large (small), we mean black holes with
nonvanishing (vanishing) classical horizon area.} The only large
${\cal N}=2$ black holes whose exact degeneracies are known are
those derived from ``compactifications" on certain {\it local}
Calabi-Yau 3-folds. Recent progress along this direction includes
\refs{\VafaQA,\AganagicJS}.

The difficulty of deducing exact degeneracies for black holes in
$\CN=2$ compactifications stems from the complexity of the
underlying Calabi-Yau. This motivates the study of
compactifications and black holes with more supersymmetry, which
typically involve simpler Calabi-Yaus. The two examples which we
will study in this paper are type IIA string theory compactified
on $K3\times T^2$, which has $\CN=4$ supersymmetry in four
dimensions; and type IIA compactified on $T^6$, which has $\CN=8$
supersymmetry in four dimensions.

The most well-studied black holes in this context are the small
$1/2$ BPS black holes in $\CN=4$ compactification on $K3\times
T^2$ \refs{\dexact,\ddmp,\DabholkarDT}. These are known as
Dabholkar-Harvey states \refs{\DabholkarJT,\DabholkarYF}, because
they are dual to wound fundamental strings in the dual heterotic
description. Their exact degeneracies are quite simple, being
given by the partition function of the heterotic string.
Unfortunately, since their classical entropy vanishes, it is not
entirely clear whether the OSV proposal should apply to these
black holes.

In order to have nonvanishing classical entropy, BPS black holes
in $\CN=4$ and $\CN=8$ string theories must preserve exactly $1/4$
and $1/8$ of the supersymmetries, respectively. Formulas for the
exact degeneracies of such black holes were recently derived in
\ssy\ and \ssyii. (The formula for the exact degeneracies of $1/4$
BPS black holes was conjectured by Dijkgraaf, Verlinde and
Verlinde nearly 10 years ago in \DVV.)

These exact degeneracies have all been tested against the
topological string amplitudes in some way or another
\refs{\dexact,\ddmp,\DabholkarDT,\CardosoXF,\PiolineVI}. In all
these tests, however, the approach has been to either Legendre or
Laplace transform the topological string amplitudes and compare
them with the black hole entropy. Only partial agreement has been
found this way, in part because of various ambiguities, e.g.\ in
the contour and measure of integration, inherent in this approach.

In this paper we will adopt a complementary viewpoint - we will
directly compare the semi-canonical black hole partition function
in ${\cal N}=4$ and ${\cal N}=8$ string theories with the
topological string partition function. This approach has several
advantages. Since it involves a sum over charges instead of an
integral over potentials, it avoids the ambiguities regarding the
contour and measure of integration. Moreover, the direct approach
can, in principle, provide a non-perturbative completion of the
topological string amplitudes.

Regardless of how one performs the comparison with the topological
string, there is always one ambiguity of the OSV conjecture in the
context of ${\cal N}=4$ and ${\cal N}=8$ string theories. This
comes from the extra charges associated with the KK gauge fields.
These charges are not present in $\CN=2$
compactification.\foot{One way to see this is to note that in an
$\CN=2$ reduction of the full supersymmetry, these gauge fields
belong to $\CN=2$ gravitini multiplets, which do not arise in
$\CN=2$ compactification.} Since the original proposal of OSV was
in the context of $\CN=2$ compactification, it is unclear whether
we should sum over these extra charges when computing the
semi-canonical partition function \zzs. On the other hand, there
is good reason to suppose that the attractor mechanism applies
just as well to these extra charges (see e.g.\FerraraUM).

Perhaps the most straightforward prescription is to turn off these
extra charges and compare the ``reduced" OSV transform of
$\Omega(p,q)$ with a similarly reduced topological string
partition function. The other natural prescription is to perform a
``full" OSV transform of the exact degeneracies with respect to
all the charges. To parameterize our ignorance, we will introduce
an integer $n$ and consider the most general OSV transform
\eqn\osvmostgenintro{ Z^{(n)}_{BH}(p,\phi) \equiv
\sum_{q_1,\dots,q_n}\sum_{q_a}\Omega_{BH}(p,q)e^{-q\cdot\phi} }
where $q_a$ denote the charges in $\CN=2$ graviton and vector
multiplets, and $q_1,\dots,q_n$ denote the extra charges
associated with the $\CN=2$ gravitini multiplets. Here
$\Omega_{BH}(p,q)$ is the proper supersymmetric index of the black
holes as defined in \refs{\GaiottoGF,\ssy,\ssyii}. Our goal for
the rest of this paper will be to calculate, and then to interpret
\osvmostgenintro.

For various technical reasons, we will restrict ourselves to
vanishing D6 brane charge; other than this restriction, our
results will be valid for arbitrary charge configurations.

Interestingly, we will see that the answer for any $n$ can be
expressed in terms of the topological string amplitude in simple
ways. In particular, for the ``full" OSV transform, we find
\eqn\resfor{ Z_{BH}^{(full)}(p,\phi) = \sum_{\phi\to \phi+2\pi ik}
|Z_{top}|^2 |g_{top}|^{2(b_1-2)} V_X + {\cal O}
(e^{-V_X/g_{top}^2}) } where $V_X$ is the volume of the Calabi-Yau
$X=K3\times T^2$ or $T^6$, and $b_1$ is the first Betti number of
$X$. Additionally, we find that in the $\CN=8$ case, the
nonperturbative corrections can be summed up completely, yielding
a closed-form expression for the exact OSV transform.

With a bit of guesswork, we also propose the following unified
presentation of the most general OSV transform:
\eqn\redfo{ Z_{BH}^{(n)}(p,\phi) = \sum_{\phi\to \phi+2\pi ik}
|Z'_{top}|^2 \sqrt{\det\,g^{(q)}}+non~pert.
 }
where $|Z_{top}'|^2$ is the square of the topological string
partition function {\it including the holomorphic anomaly}, and
$g^{(q)}$ is a ``quantum-corrected" metric on the moduli space of
the topological string on $X$ (to be defined below).

This paper is organized as follows. In section 2 we study the
$1/8$ BPS black holes in ${\cal N}=8$ string theory; compute the
semi-canonical partition function $Z_{BH}(p,\phi)$ including the
nonperturbative corrections; and compare to the topological
string. In section 3 we shall study the analogous problem for
$1/4$ BPS black holes in ${\cal N}=4$ string theory, although the
analysis of the nonperturbative corrections will be left to
appendix A. Finally, section 4 summarizes our results and
discusses the possible implications.

\newsec{$\CN=8$ Black Holes}

\subsec{Preliminaries}

In this section, we wish to consider the OSV transform of the exact
degeneracies of 1/8 BPS $\CN=8$ black holes. These black holes can
be realized as wrapped branes and strings in a type IIA
compactification on $T^6$. A formula for their exact degeneracies
was derived recently in \ssyii. It takes the form \eqn\degenne{
\Omega_{\CN=8}(p,q) \equiv \sum_{J^3, BPS~states} 2(J^3)^2
(-)^{2J^3} = d(J(p,q)) = \oint d\rho\, F(\rho)e^{-2\pi i \rho
J(p,q)} } where the two ingredients of the RHS are:

 \lfm{1.} $F(\rho)$, a modular form
\eqn\Frhodef{ F(\rho) = {\theta_3(2\rho)\over \eta(4\rho)^6} =
{\eta(2\rho)^{5}\over \eta(4\rho)^{8}\eta(\rho)^{2}}\sim C\rho^{5/2}
e^{\pi i/8\rho}\left(1+\dots\right) } The $\dots$ represents an
expansion in powers of $e^{-\pi i/2\rho}$. These corrections
contribute order ${\cal O}(e^{-Q^2})$ terms to $\Omega_{\CN=8}$,
which are expected to be non-perturbative from the topological
string point of view.

 \lfm{2.}$J(p,q)$, the unique quartic invariant of the $\CN=8$
U-duality group, known as the Cremmer-Julia invariant \CremmerUP.
In a suitable basis (see e.g.\ \refs{\vijay,\KalloshUY}), this
invariant takes the form
\eqn\CJlit{ J = -\Tr(YZYZ)+{1\over4}(\Tr YZ)^2-4({\rm Pf} Y+{\rm
Pf} Z) } where $Y$ and $Z$ are $8\times8$ antisymmetric real
matrices which encode the charges of wrapped branes and strings.

\bigskip

From the ${\cal N}=8$ point of view $Y$ and $Z$ are naturally
electric and magnetic charges. However, in performing the OSV
transform, we should assign the magnetic and electric charges that
are consistent with the coupling of ${\cal N}=2$ vector, graviton
and gravitini multiplets, as follows:
 \eqn\yzdef{ Y=\left(\matrix{ P & -p^1 & -p^2 \cr p^{1T} & 0 &-q_0 \cr p^{2T}& q_0 & 0}\right),\qquad
 Z=\left(\matrix{Q & -q_1 & -q_2 \cr q_1^T & 0 &-p^0\cr q_2^T & p^0 & 0}\right)
 }
Here $P$ and $Q$ are $6\times 6$ antisymmetric matrices, and
$q_1$, $q_2$, $p^1$, $p^2$ are six-dimensional vectors. These are
divided into electric charges $q_\Lambda$ and magnetic charges
$p^\Lambda$, $\Lambda=0,\dots,27$, as follows:
\eqn\pqdefne{
 q_\Lambda = \{q_0,~Q_{ij},~q_{1i},~q_{2i}\},\qquad p^\Lambda =
 \{p^0,~P^{ij},~p^{1i},~p^{2i}\}
 }
Physically, $P^{ij}$ and $Q_{ij}$ are identified with
$T^6$-wrapped D4 and D2 brane charges respectively, with
$i,j=1,\dots,6$ representing the 1-cycles of the $T^6$. Meanwhile,
$q_{1i}$ and $q_{2i}$ correspond to KK monopole and NS5 brane
charges, respectively, while $p^{1i}$ and $p^{2i}$ correspond to
KK momentum and F-string winding charges, respectively. Finally,
$q_0$ is D0 brane charge, while $p^0$ is D6 brane charge.

As mentioned in the introduction, we will focus exclusively on the
case $p^0=0$ in this paper. In this limit, one can check that
\CJlit\ simplifies to
\eqn\jsimd{ J = 4(p)^3\left(q_0 +{1\over 12} D^{AB}q_Aq_B\right) }
where $A,B=1,\dots,27$ are the collective indices for the 27
electric and magnetic charges other than $q_0$ or $p^0$, and we
have defined \eqn\dabc{ (p)^3= D_{ABC}p^Ap^Bp^C = {\rm Pf}P+p^{1T}
P p^2,~~~~D_{AB}=D_{ABC}p^C,~~~~ D^{AB}D_{BC}=\delta^A_C.
 }
This result illustrates the advantages of assuming $p^0=0$, in the
following way. Notice how $J\sim q^2$ for $p^0=0$. On the other
hand, for $p^0\ne 0$, one can show that $J\sim p^0 q^3$ for large
electric charges $q$. Since the black hole degeneracy grows like
$\Omega_{BH}(p,q) \sim \exp\left(\pi\sqrt{J(p,q)}\right)$ in the
large charge limit, convergence of the OSV transform
\osvmostgenintro\ seems problematic for $p^0\ne 0$, while it is
less so for $p^0=0$. Thus, in addition to simplifying the
calculations considerably, the restriction $p^0=0$ also improves
the convergence of the OSV transform \osvmostgenintro.

Now that we have described the various components of the exact
degeneracy formula \degenne, let us turn to evaluating the OSV
transform
\eqn\osvtransfne{
Z_{\CN=8}(p,\phi) =
\sum_{q_0,q_A}\Omega_{\CN=8}(p,q)e^{-q_0\phi^0-q_A\phi^A}
}

\subsec{The full OSV transform}

Let us start by rewriting the black hole partition sum
\eqn\bhsuman{\eqalign{ Z_{\CN=8}(p,\phi)&=\sum_{q_0,q_A} \Omega_{\CN=8}(p,q)
e^{-q_0\phi^0-q_A\phi^A} \cr &= {1\over
(p)^{3}}\sum_{\phi^0\to\phi^0+2\pi ik^0}\sum_{q_A} \sum_J d(J)
\exp\left[ -\phi^0\left({J\over 4(p)^3}-{1\over
12}D^{AB}q_Aq_B\right)-q_A\phi^A \right]
 }}
where we used \jsimd\ to express $q_0$ in terms of $J$, and the
sum over $k^0$ is in the range $0\leq k^0\le 4(p)^3-1$. The sum
over $J$ gives precisely the modular form $F(i\phi^0/8\pi (p)^3)$
defined in \Frhodef. Meanwhile, we can evaluate the sum over $q_A$
via Poisson resummation.\foot{Due to the non-positive-definiteness
of $D_{AB}$, the sum over $q_A$ is, strictly speaking, not
convergent. We will regularize the divergence by performing a
formal Poisson resummation. A more careful regularization was
considered in section 6 of \DabholkarDT. In any event, we expect
our results to be regularization independent.} Then we are left
with\foot{Here and throughout we will neglect overall numerical
factors.}
\eqn\bhsab{\eqalign{ &Z_{\CN=8}(p,\phi)=\sum_{\phi^0\to\phi^0+2\pi ik^0}\sum_{\phi^A\to \phi^A+2\pi
ik^A} (\phi^0)^{-{27\over 2}}(p)^{21\over2} F\left({i\phi^0 \over
8 \pi(p)^3}\right) \exp\left( -{3D_{ABC}\phi^A\phi^Bp^C\over
\phi^0} \right)}} Finally, we can make use of the modular property
of $F(\rho)$,
\eqn\fftti{ F(\rho) = \rho^{5/2} e^{\pi i\over 8\rho} \tilde
F(-1/4\rho)} with
\eqn\Ftildedef{
\tilde F(\rho) = e^{\pi i\rho/2}{\theta_3(2\rho)\over
\eta(\rho)^6} = 1+\CO\left(e^{2\pi i\rho}\right) } and write
\bhsab\ as
 \eqn\bhsacn{ Z_{\CN=8}(p,\phi)=\sum_{\phi\to \phi+2\pi ik}(\phi^0)^{-11}
(p)^3\tilde F\left( {2\pi i(p)^3\over\phi^0} \right) \exp\left(
{\pi^2 (p)^3\over\phi^0}-{3D_{ABC}\phi^A\phi^Bp^C\over \phi^0}
\right)} In the next subsection, we will see how this fairly
simple expression becomes even simpler when it is recast in terms
of the topological string on $T^6$.

\subsec{Rewriting in terms of the topological string on $T^6$}

The topological string on $T^6$ is essentially trivial. The
prepotential consists of only a tree-level term, determined by the
intersection form on $T^6$:
\eqn\FtopNeight{ Z_{top}=e^{F_{top}},\qquad F_{top} ={(2\pi i)^3D_{ABC}t^A t^B t^C\over g_{top}^2}
}
Here $t^A$ are the K\"ahler moduli of $T^6$. Together with the
topological string coupling $g_{top}$, they are related to the
magnetic charges and electric potentials defined above via the
attractor equations:
\eqn\attractorne{
t^A = {X^{A}\over X^0} = {p^{A}+i\phi^A/\pi\over
p^0+i\phi^0/\pi},\qquad g_{top} = {4\pi i \over X^0} = {4\pi i
\over p^0+i\phi^0/\pi}
}
Keep in mind that we are considering only the case $p^0=0$ in this
work. Notice also that we have slightly generalized the notion of
$F_{top}$ to include the 12 moduli in the $\CN=2$ gravitini
multiplets, $t^{1i}$ and $t^{2i}$. This is well-motivated from
U-duality, and we will see that it leads the correct answer.

Another quantity of interest is the K\"ahler potential of the
underlying special geometry (the factor of $1/\pi$ is for later
convenience):
 \eqn\Kahlerdef{ e^{-K} =
{1\over\pi}\sum_{\Lambda=0}^{27}{\rm Re}\,\bar
X^{\Lambda}\partial_{\Lambda}F_{top}
 }
For $p^0=0$, this is simply
 \eqn\Kahlernepz{
 e^{-K} = {2\pi(p)^3\over \phi^0}
}

Now we are ready to recast the OSV transform in terms of the
topological string on $T^6$. Using \FtopNeight--\Kahlernepz, we
see that \bhsacn\ can be written as
\eqn\intfull{\eqalign{ Z_{\CN=8}
 &=\sum_{\phi\to\phi+2\pi ik}|Z_{top}|^2 |g_{top}|^{10}e^{-K}
 \tilde F\left(ie^{-K}\right)
 }}
We stress that \intfull\ is only derived in the case $p^0=0$. We
also note that the sum over $k^0$ is restricted to the range
$0\leq k^0<4(p)^3$, although $(p)^3$ is large in the limit of
small $g_{top}$ while fixing the size of the $T^6$.

To highlight the $g_{top}$ dependence, it is useful to rewrite
\intfull\ in terms of the volume of the $T^6$:
\eqn\volre{ V_{T^6}=|g_{top}|^2 e^{-K} } Then \intfull\ can be
rewritten as
\eqn\intfuls{ Z_{\CN=8} = \sum_{\phi\to
\phi+2\pi ik} |Z_{top}|^2 |g_{top}|^8 V_{T^6} \tilde
F\left({iV_{T^6}\over |g_{top}|^2}\right)
}
>From \Ftildedef, we see that this is essentially $|Z_{top}|^2$ up
to corrections that are nonperturbative in $g_{top}$,
 \eqn\intfulsexpand{ Z_{\CN=8}=\sum_{\phi\to \phi+2\pi ik} |Z_{top}|^2 |g_{top}|^8 V_{T^6}
 +nonpert.
 }
Interestingly, the nonperturbative corrections go like $\sim
e^{-2\pi V_{T^6}/|g_{top}|^2}$. We will discuss possible
interpretations of this in section 4.

\subsec{The general OSV transform}

Before we go on to the case of $1/4$ BPS $\CN=4$ black holes, let
us take a moment to consider a ``reduced" OSV transform which is
natural from the point of view of $\CN=2$ supersymmetry. Here we
transform $Z_{\CN=8}(p,q)$ only with respect to the the charges
$q_0$ and $Q_{ij}$ ($i,j=1,\cdots,6$) which are associated to
$\CN=2$ vector multiplets. The extra magnetic charges $p^1$, $p^2$
and electric potentials $\phi^1$, $\phi^2$ in $\CN=2$ gravitini
multiplets are turned off. The calculation of this ``reduced" BH
partition function is almost identical to the previous section,
and the result is
\eqn\redne{ \eqalign{ Z_{\CN=8}^{red} &= \sum_{\phi\to \phi+2\pi
i k} |Z_{top}|^2 |g_{top}|^6 e^{+K} \tilde F\left(ie^{-K}\right)
\cr &= \sum_{\phi\to \phi+2\pi i k} |Z_{top}|^2 {|g_{top}|^8}{1
\over V_{T^6}
}  + nonpert.
 }}

For the sake of completeness, let us also compute the most general
``reduced" partition function, with $0\le n\le 12$ of the
gravitini charges summed up:
\eqn\halfcano{\eqalign{ Z_{\CN=8}^{(n)} &=  \sum_{\phi\to \phi+2\pi i k}
|Z_{top}|^2 |g_{top}|^8 (V_{T^6})^{n/6-1}\, \tilde
F\left(ie^{-K}\right) \cr
 &= \sum_{\phi\to \phi+2\pi i k}
|Z_{top}|^2 |g_{top}|^8(V_{T^6})^{n/6-1}+nonpert.
 }}
So, for instance, $n=0$ corresponds to \redne, while $n=12$
corresponds to \intfuls--\intfulsexpand. It is amusing to note
that the ``half-reduced" transform with $n=6$ gives the simplest
result, essentially $|Z_{top}|^2$ with no volume factor.

\newsec{$\CN=4$ Black Holes}

\subsec{Preliminaries}

Having computed the OSV transform of the exact $\CN=8$
degeneracies, let us now consider the analogous problem for the
exact degeneracies of 1/4 BPS $\CN=4$ black holes. We will
consider type IIA string theory compactified on $K3\times T^2$.
The BPS black holes can then be described as supersymmetric bound
states of branes and strings wrapped on various cycles in
$K3\times T^2$. A formula for their exact degeneracy was
conjectured in \DVV\ and derived in \ssy. It takes the
form\foot{Note the extra factor of $(-1)^{q_e\cdot q_m}$ relative
to \DVV. This factor was missed previously, and it is a
consequence of the 4D-5D correspondence \refs{\GaiottoGF,\ssy}. We
will see below that this factor is essential in order to obtain
the correct OSV transform.
}
\eqn\DVVdegen{ \Omega_{\CN=4}(p,q)\equiv \sum_{J^3,BPS~states}
(-)^{2J^3}=d(q_e^2,q_m^2,q_e\cdot q_m) = \oint d\rho d\sigma d\nu
{e^{\pi i(\rho q_m^2+\sigma q_e^2+(2\nu-1) q_e\cdot q_m)} \over
\Phi(\rho,\sigma,\nu)} } where the ingredients of the RHS are

\lfm{1.} $\Phi(\rho,\sigma,\nu)$, the unique weight 10 automorphic
form of the modular group $Sp(2,{\Bbb Z})$. Although quite
complicated in general, $\Phi$ simplifies near ``rational
quadratic divisors" (RQDs), along which $1/\Phi$ has double poles.
For details about the approximation of $\Phi$ by rational
quadratic divisors, see \DVV\ and also section 3 of \CardosoXF.
The divisor whose contribution dominates the asymptotic degeneracy
is \eqn\ratl{ \rho\sigma+\nu - \nu^2=0 } The contributions of the
other divisors are suppressed by ${\cal O}(e^{-Q^2})$, which is
expected to be non-perturbative from the point of view of
topological strings. (In appendix A, we study the effect of these
subleading divisors and verify that they are indeed
nonperturbative.) Working perturbatively in $1/Q$, we can make the
approximation
\eqn\pertdvv{ {1\over \Phi(\rho,\sigma,\nu)} =
 {\sigma^{12}\over (\rho\sigma+\nu-\nu^2)^2}
 \eta\left({\rho\sigma-\nu^2\over\sigma}\right)^{-24}\eta\left(-{\rho\over
\rho\sigma-\nu^2}\right)^{-24} +
\CO\left((\rho\sigma+\nu-\nu^2)^0\right)
}

\lfm{2.} $q_e^2$, $q_m^2$ and $q_e\cdot q_m$, invariants of the
$SO(6,22;{\Bbb Z})$ subgroup of the full $\CN=4$ U-duality group
$SL(2,{\Bbb Z})\times SO(6,22;{\Bbb Z})$. These invariants encode
the charges of various branes and strings wrapped on the cycles of
$K3\times T^2$. Explicitly, we have
\eqn\dualityinvacnf{\eqalign{
&q_e^2=2q_0p^1+C^{MN}q_Mq_N\cr &q_m^2=2p^0q_1+C_{MN}p^Mp^N\cr
&q_e\cdot q_m = p^0q_0+p^1q_1-p^Mq_N }} where $M,N=2,\dots,27$,
and
\eqn\CMNdef{ C_{MN}x^My^N =
C_{ab}x^ay^b+x^{24}y^{27}+x^{25}y^{26}+x^{26}y^{25}+x^{27}y^{24}
}
with $C_{ab}$, $a=2,\dots,23$ the intersection form on $H^2(K3)$.
Here the electric charges are given by the following: $q_0$ is
D0-charge; $q_1$ is $T^2$-wrapped D2 charge; $q_a, a=2,...23$ is
$K3$-wrapped D2 charge; and $q_i, i=24,...27$ are momentum and
winding modes of $K3\times S^1$-wrapped NS5 branes. Meanwhile, the
magnetic charges correspond to the following: $p^0$ is D6-charge;
$p^1$ is $K3$-wrapped D4 charge; $p^a$ are $T^2\times ({\rm K3
~cycle})$-wrapped D4 charge;  and $p^i$ are F-string $T^2$
momentum/winding modes.

\subsec{The full OSV transform}

Just as in the $\CN=8$ case, we shall restrict to the case $p^0=0$
which greatly simplifies the calculations. We will use the
shorthand notation
\eqn\shorthand{
Q=q_e^2,\qquad P=q_m^2,\qquad R=q_e\cdot q_m
}
The black hole partition sum can be written
\eqn\nfcas{ \eqalign{
 Z_{\CN=4}(p,\phi)&= \sum_{q_0,q_1,q_M}
d(Q,P,R) e^{-q_0\phi^0-q_1\phi^1-q_M\phi^M} \cr &= (p^1)^{-2}
\sum_{q_M} \sum_{\phi^{0,1}\to \phi^{0,1}+2\pi ik^{0,1}}
\sum_{Q,R} d(Q,P,R)\cr &~~~~~~~~~\times \exp\left[ -{\phi^0\over
2p^1}(Q-C^{MN}q_Mq_N) - {\phi^1\over p^1} (R+p^Mq_M)-q_M\phi^M
\right]}
 }
where $k^{0,1}$ are summed in the range $0,\cdots,p^1-1$ (note
that $Q$ is even). The sum over $Q$ and $R$ yields (by definition)
\eqn\QRsum{
 \sum_{Q,R}d(Q,P,R)e^{ -{\phi^0\over
2p^1}Q - {\phi^1\over p^1} R}=\oint d\rho\, {e^{\pi i\rho P}\over
\Phi(\rho,\sigma={\phi^0\over 2\pi ip^1},\nu = {\phi^1\over 2\pi
ip^1}+\half)}
}
The idea is to compute the contour integral in \QRsum\ in terms of
the residues at the relevant RQDs. As discussed above, the
dominant contribution to the integral is expected to be given by
the RQD \ratl--\pertdvv. Integrating $\rho$ around this RQD yields
\eqn\domdiv{\eqalign{ \oint d\rho\, {e^{\pi i\rho P}\over
\Phi(\rho,\sigma_*,\nu_*)} &= \sigma^{10}\partial_\rho \left[
{e^{\pi i\rho P}\over \eta({\rho\sigma-\nu^2\over \sigma})^{24}
\eta(-{\rho\over \rho\sigma-\nu^2})^{24} }
\right]_{\rho=\rho_*,~\sigma=\sigma_*,~\nu=\nu_*}+\dots\cr &=
{\sigma_*^{10} e^{\pi i\rho_* P}\over \eta(-{\nu_*\over
\sigma_*})^{24} \eta({\rho_*\over\nu_*})^{24}
}  \left[\pi iP - 24 {\eta'(-{\nu_*\over
\sigma_*})\over \eta(-{\nu_*\over \sigma_*})}-24
{\eta'({\rho_*\over\nu_*})\over \eta({\rho_*\over\nu_*})} \right]
+ \dots }} where we have defined
\eqn\defsss{ \eqalign{ &\sigma_*={\phi^0\over 2\pi ip^1},~~~~\nu_*
= {\phi^1\over 2\pi ip^1}+\half,~~~~\rho_* = {\nu_*^2-\nu_*\over
\sigma_*} = {(\pi p^1)^2+(\phi^1)^2\over 2\pi i p^1\phi^0}, \cr
 } }
and $\dots$ refers to the contributions of the other RQDs. In
appendix A, we study these contributions, and we show that they
are indeed nonperturbative in $g_{top}$ relative to the dominant
contribution \domdiv.

This leaves the sum over $q_M$, which can be evaluated using a
(formal) Poisson resummation (as for $\CN=8$, there are issues of
convergence we are glossing over, since $C^{MN}$ is not
positive-definite):
\eqn\nfpoisson{
\sum_{q_M}e^{{\phi^0\over 2p^1}C^{MN}q_Mq_N - ({\phi^1\over
p^1}p^M+\phi^M) q_M} =
\left({p^1\over\phi^0}\right)^{13}\sum_{\phi^M\to\phi^M+2\pi i
k^M}e^{-{ C_{MN} (p^1\phi^M+{\phi^1 p^M}) (p^1\phi^N+\phi^1
p^N)\over 2p^1\phi^0}}
}
Combining this with \domdiv\ and \defsss, \nfcas\ becomes after
some algebra
\eqn\zfnewa{ \eqalign{ Z_{\CN=4}&=
\sum_{\phi\to \phi+2\pi ik} {p^1\over(\phi^0)^3} {
\exp\left({{C_{MN}(\pi^2 p^1p^M p^N-2\phi^1p^M
\phi^N-p^1\phi^M\phi^N)\over2\phi^0}}\right) \over\eta({\pi
p^1-i\phi^1\over i\phi^0})^{24} \eta({\pi p^1+i\phi^1\over
i\phi^0})^{24}
}  \cr
 &\qquad\qquad\qquad\qquad\times\left[\pi i C_{MN}p^M p^N - 24 {
\eta'({\pi p^1-i\phi^1\over i\phi^0}) \over \eta({\pi
p^1-i\phi^1\over i\phi^0})}-24 {\eta'({\pi p^1+i\phi^1\over
i\phi^0})\over \eta({\pi p^1+i\phi^1\over i\phi^0})} \right]+\dots
}}
In the next subsection, we will see how this rather complicated
result of the OSV transform has an extremely simple interpretation
in terms of the topological string on $K3\times T^2$.

\subsec{Rewriting in terms of the topological string on $K3\times
T^2$}

The topological string on $K3\times T^2$ is only slightly less
trivial than the topological string on $T^6$. The prepotential
consists of a tree-level term together with a one-loop correction
coming from worldsheet instantons:
\eqn\FtopNfour{ Z_{top}=e^{F_{top}},\qquad F_{top} =
 {(2\pi i)^3C_{MN}t^M t^N t^1\over 2g_{top}^2}-24\log\eta(t^1)
}
Here $t^1$, $t^M$, $t^N$, $M,N=2,\dots,27$ are the K\"ahler moduli
of $K3\times T^2$. In particular, $t^1$ corresponds to the
complexified K\"ahler modulus of $T^2$. Just as for $\CN=8$, we
have generalized the notion of $F_{top}$ to include the 4 moduli
in the $\CN=2$ gravitini multiplets, $t^{24,\dots,27}$. Again,
this is well-motivated from U-duality, and it correctly reproduces
the result of the OSV transform.

The attractor equations work exactly the same as for $\CN=8$
\attractorne. Explicitly, we have (recall we are setting $p^0=0$)
\eqn\attractornf{
t^M = {\pi p^{M}+i\phi^M\over i\phi^0},\qquad t^1 = {\pi
p^{1}+i\phi^1\over i\phi^0} \qquad g_{top}  = {4\pi^2 \over
\phi^0}
}

Finally, we find the K\"ahler potential in this case to be:
 \eqn\Kahlernfpz{
 e^{-K} = -{2ip^1\over\phi^0} \left[ \pi i C_{MN}p^M p^N-{24\eta'({\pi p^1-i\phi^1\over i\phi^0})
 \over \eta({\pi
p^1-i\phi^1\over i\phi^0})}- {24\eta'({\pi p^1+i\phi^1\over
i\phi^0})\over \eta({\pi p^1+i\phi^1\over i\phi^0})}\right] } Notice
that we have generalized slightly the definition of the K\"ahler
potential to include the one-loop topological string amplitude (the
second and third terms of \Kahlernfpz, due to world sheet
instantons). Normally, $e^{-K}$ is computed using the tree-level
prepotential alone. Thus, we can think of $e^{-K}$ defined in
\Kahlernfpz\ as a ``quantum-corrected" K\"ahler potential.

Now we have all the ingredients necessary to interpret the OSV
transform \zfnewa\ in terms of the topological string on $K3\times
T^2$. In fact, using \FtopNfour--\Kahlernfpz, the complicated
result \zfnewa\ simplifies considerably:
\eqn\intfullnf{\eqalign{ Z_{\CN=4}
 &=\sum_{\phi\to\phi+2\pi ik}|Z_{top}|^2 |g_{top}|^{2}e^{-K}
 +\dots
 }}
Again, keep in mind that we have made the crucial restriction
$p^0=0$ in deriving \intfullnf. We also recall from \nfcas\ that
the sum over $\phi^0$ and $\phi^1$ shifts only runs over the range
$0\leq k^{0},~k^1 < p^1$.

Once again, we can rewrite the answer in terms of the
(quantum-corrected) volume of $V_{K3\times T^2}$:
\eqn\volkthreettwo{ V_{K3\times T^2}=|g_{top}|^2 e^{-K} }
Then \intfullnf\ becomes even simpler,
\eqn\intfulsnf{ Z_{\CN=4} = \sum_{\phi\to
\phi+2\pi ik} |Z_{top}|^2 V_{K3\times T^2} +\dots
}
This should be compared with the analogous $\CN=8$ result
\intfulsexpand. Aside from the factors of $g_{top}$ appearing in
the latter, which could perhaps be absorbed into the definition of
$Z_{top}$, we see that the two results are in complete agreement.

\subsec{The general OSV transform}

Once again we can transform the black hole degeneracy only with
respect to only $q_0$ and the charges $q_{a=1,\dots,23}$ which are
associated to $\CN=2$ vector multiplets. The calculation is nearly
identical to the full transform discussed above; the only
difference is the power of $p^1/\phi^0$ in \nfpoisson\ becomes 11
instead of 13. The resulting ``reduced" BH partition function is
\eqn\rednff{ Z_{\CN=4}^{red} = \sum_{\phi\to\phi+2\pi
ik} |Z_{top}|^2 {V_{K3}\over V_{T^2}} +\cdots } where we have
defined $V_{T^2}=|{\rm Im}\,t^1| = {\pi p^1\over\phi^0}$ and
$V_{K3}={V_{K3\times T^2}\over V_{T^2}}$.

Finally, let us also consider the most general ``reduced"
partition function obtained by summing up $0\le n\le 4$ gravitini
charges. The result is:
\eqn\halffna{ Z_{\CN=4}^{(n)} =\sum_{\phi\to\phi+2\pi ik}
|Z_{top}|^2 V_{K3}(V_{T^2})^{n/2-1} +\cdots }

\newsec{Summary and Discussion}

\subsec{Summary of results}

In the previous sections we have evaluated the OSV transform for
$1/4$ and $1/8$ BPS black holes in type IIA compactification on
$K3\times T^2$ and $T^6$, respectively. Apart from the requirement
of vanishing D6 brane charge ($p^0=0$), the black holes we
considered had completely general charge configurations. In
particular, they could either include the charges in ${\cal N}=2$
gravitini multiplets, or not.

In all cases, we found that the results take the form
\eqn\OSVtransfgen{ Z_{BH}^{(n)}(p,\phi) \equiv
\sum_{q}\Omega_{BH}(p,q)e^{-q\phi}=
\sum_{k\in\Gamma}G^{(n)}(p,\phi+2\pi i k)
} with
\eqn\Gdef{
G^{(n)}(p,\phi)=|Z_{top}|^2\times ({\rm simple~factor})
}
where $n$ denotes the number of gravitini charges being summed
over ($n=4$ and $n=12$ for the full $\CN=4$ and $\CN=8$
transforms, respectively), and the precise form of the simple
factor depends on the transform and degeneracy being considered.

In more detail, we found that up to nonperturbative corrections in
$g_{top}$, the answer for $G$ was
\eqn\OSVtransflistii{\eqalign{
  & G_{\CN=8}^{(n)}(p,\phi) =|Z_{top}|^2 |g_{top}|^8(V_{T^6})^{n/6-1}   \cr &
G_{\CN=4}^{(n)}(p,\phi) =  |Z_{top}|^2 V_{K3} (V_{T^2})^{n/2-1}
} }
In addition, we were able to sum up the nonperturbative
corrections to the $\CN=8$ answer, yielding an exact result in
this case:
\eqn\fneosv{ G_{\CN=8}^{(n)}(p,\phi) = |Z_{top}|^2 |g_{top}|^8
(V_{T^6})^{n/6-1} \tilde F\left({iV_{T^6}\over |g_{top}|^2}\right)
 }
with the modular form $\tilde F(\rho)$ given by \Ftildedef.

Despite the discrepancies from $|Z_{top}|^2$, the OSV transforms
of the ${\cal N}=4$ and ${\cal N}=8$ degeneracies have very
similar structures. Clearly, there are patterns here that need to
be better understood!

\subsec{The holomorphic anomaly and the metric on moduli space}

Let us now examine the most general OSV transforms \halfcano\ and
\halffna\ in more detail. We can write them in terms of a natural
metric on the space of $X^\Lambda$'s - the one that governs the
kinetic term of the corresponding vector fields in supergravity:
\eqn\tauijdef{ g_{\Lambda\Sigma}^{(cl)} \equiv
\partial_{\Lambda}\partial_{\bar\Sigma}e^{-K^{(cl)}} =
{1\over\pi}{\rm
Re}\,\partial_{\Lambda}\partial_{\Sigma}F_{top}^{(cl)}
 }
where the $(cl)$ superscript denotes the classical tree-level
contribution; and $\Lambda,\Sigma=0,\dots,23+n$ for $K3\times T^2$
and $\Lambda,\Sigma=0,\dots,15+n$ for $T^6$. A short calculation
using the appropriate $F_{top}^{(cl)}$ shows that the determinant
of $g^{(cl)}$ takes the form (disregarding overall numerical
factors)
\eqn\detimtau{\eqalign{
 \det\, g^{(cl)}&= ({\rm Im}\,t^1)^{20+n}(V_{K3\times
T^2}^{(cl)})^2\qquad\quad (K3\times T^2)\cr \det\, g^{(cl)} &=
(V_{T^6})^{6+n/3} \qquad\quad\quad\qquad\qquad\,\,\, (T^6) }}
Therefore, the general OSV transforms \halfcano\ and \halffna\ can
be written nicely in terms of this metric as
\eqn\OSVtransflistrew{\eqalign{
 G^{(n)}(p,\phi) &= |Z_{top}'|^2\sqrt{\det\,g^{(cl)}}
 \left({V_{X}\over V_{X}^{(cl)}}\right)\cr
}}
where $X=K3\times T^2$ or $T^6$, and
\eqn\Ztopprimdef{
|Z_{top}'|^2 = \cases{ |Z_{top}|^2 ({\rm Im}\,t^1)^{-12} &
$X=K3\times T^2$\cr |Z_{top}|^2 e^{4K} & $X=T^6$}
 }
Notice how the explicit $n$ dependence has been completely
absorbed into the determinant of the metric. In particular,
$|Z_{top}'|^2$ is independent of $n$. In the case of $K3\times
T^2$, we recognize $|Z_{top}'|^2$ to be the square of the
topological string partition function {\it including the
holomorphic anomaly} \HarveyIR!

The factor $e^{4K}$ in the $\CN=8$ case is more mysterious, since
the topological string on $T^6$ has no holomorphic anomaly. Let us
consider the special case when only the K\"ahler moduli
$t^1,t^2,t^3$ are turned on, and then
\eqn\efk{e^{4K}=|g_{top}|^8({\rm Im} t^1 {\rm Im}t^2 {\rm
Im}t^3)^{-4}} Apart from the factor $|g_{top}|^8$, which also showed
up in \OSVtransflistii, the RHS of \efk\ is identical to the 1-loop
holomorphic anomaly in toroidal orbifold models, such as $T^6/{\Bbb
Z}_3\times {\Bbb Z}_3$. We suspect that \efk\ may be interpreted as
the holomorphic anomaly of a modified genus 1 amplitude defined by a
new index. It would be interesting to understand this.

Although \OSVtransflistrew\ is fairly nice overall, it does
contain the rather ugly factor ${V_{X}\over V^{(cl)}_{X}}$. We
will attempt to explain this factor as follows. It differs from
unity only in the case of $X=K3\times T^2$, so let us focus on
that case. From the form of $\det\,g^{(cl)}$ in \detimtau, it is
tempting to propose the existence of a ``quantum-corrected" metric
on the space of $X^\Lambda$'s, $g^{(q)}$, whose determinant gives
the quantum-corrected volume, i.e. \eqn\quantumdetimdau{
\sqrt{\det\,g^{(q)}}=({\rm Im}\,t^1)^{10+n/2}V_{K3\times T^2}
 }
Then in terms of this metric, \OSVtransflistrew\ reduces to
\eqn\neatlyas{ G^{(n)}(p,\phi) = |Z_{top}'|^2  \sqrt{\det\,g^{(q)}}
 }
and the calculation of the OSV transform \OSVtransfgen\ becomes
\eqn\neatlyasii{
 Z_{BH}^{(n)}(p,\phi) = \sum_{\phi\to \phi+2\pi i k}
 |Z_{top}'|^2\sqrt{\det\,g^{(q)}}
}
where again, all of the explicit $n$ dependence is contained in
the determinant of the metric. Note that \neatlyas--\neatlyasii\
also apply to $X=T^6$, since in that case it is natural to suppose
that $g^{(q)}=g^{(cl)}$.

Introducing a quantum-corrected metric $g^{(q)}$ may seem ad hoc,
but evidence for its existence comes from the following
observation. If we suppose that the quantum corrections to the
metric are somehow the result of the one-loop worldsheet instanton
corrections in \FtopNfour, then it is natural to assume that
\eqn\deltagdef{
\delta g_{\Lambda\Sigma} =
g^{(q)}_{\Lambda\Sigma}-g^{(cl)}_{\Lambda\Sigma}
}
only has nonzero components for $\Lambda,\Sigma=0,1$ and is a
function only of $X^0$ and $X^1$. Imposing that $\delta g$ be
real, we find that the nonzero components of $\delta g^{(q)}$ are
given {\it uniquely} by
 \eqn\ctaus{
\left( \matrix{ \delta g_{00} &\delta g_{01} \cr \delta g_{10}
&\delta g_{11} } \right) = {24 |g_{top}|^2\over {\rm Im}t^1} {\rm
Re}\left({\eta'(t^1)\over \eta(t^1)}\right) \left( \matrix{ |t^1|^2
& -{\rm Re} t^1 \cr -{\rm Re} t^1 & 1 }\right) } The uniqueness and
the simplicity of \ctaus\ strongly suggests that the idea of a
quantum-corrected metric on the space of $X^\Lambda$'s should be
taken seriously.

To summarize, eq.\ \neatlyas--\neatlyasii\ provide a unified
description of the most general OSV transform, for both $\CN=4$
and $\CN=8$ supersymmetry, in terms of the topological string
amplitude with holomorphic anomaly and the quantum metric on the
moduli space. The latter provides a natural measure for the
wavefunction interpretation of $Z_{top}$. A similar measure factor
was proposed in \verlinde\ (see e.g.\ eq.\ (6.6) of that paper);
however, we note that our answer differs significantly from that
of \verlinde.

\subsec{D-instantons}

As discussed in \DabholkarDT\ and the introduction, the sum over
shifts $\phi\to \phi+2\pi ik$ in \OSVtransfgen\ is entirely
expected.\foot{Of course, it is still nontrivial that the summand
is essentially $|Z_{top}|^2$ times a measure factor.} After all,
the OSV transform involves a sum of $e^{-q\cdot\phi}$ over
integral charges $q$; therefore the result must be periodic in
imaginary $\phi$. However, the interpretation of the sum from the
topological string point of view is intriguing.

First, restricting to the case $p^0=0$, we have $\phi^0={4\pi^2\over
g_{top}}$. Thus, the periodicity in imaginary $\phi^0$ is
reminiscent of the periodicity in the theta angle in Yang-Mills
theory. The latter periodicity signals the quantization of instanton
charges, and so one might expect the periodicity in imaginary
$\phi^0$ to be related to the quantization of D-instanton charges.
In fact, the OSV transform \eqn\osvsum{ \sum_q \Omega(p,q)
e^{-q\cdot\phi} = \sum_q \Omega(p,q) (-)^{p\cdot q}
\exp\left[-{4\pi^2\over g_{top}}(q_0 + q_At^A)\right] } can be
interpreted as an sum over D-instanton contributions in the
topological B-model. Our result can then be thought of as a relation
between the partition function of B-branes and the perturbative
partition function of the A-model. This is reminiscent of the
topological S-duality \NekrasovJS\ although the details seem very
different.

Meanwhile, the sum over shifts in $\phi^A$ can be written as
shifts in the K\"ahler moduli
\eqn\shiftkahhere{
t^A \to t^A+{1\over2\pi i}k^A g_{top}
}
is entirely analogous to the sum in the 2DYM/topological string
correspondence \VafaQA,
\eqn\shiftkah{ t\to t+mlg_{top},~~~~\bar t\to \bar t-mlg_{top} }
(The apparent discrepancy of $2\pi i$ is because $t=2\pi i
X^1/X^0$ in \VafaQA.) This was interpreted in \VafaQA\ as a
summation over contribution from sectors with RR 2-form fluxes. It
would be nice to understand the analogous statement in our
context.

\subsec{Non-perturbative corrections}

The full result of the OSV transform for the $\CN=8$ black hole
\intfuls\ can be thought of as the non-perturbative completion of
the topological string amplitude on $T^6$. Let us briefly discuss
the nature of the non-perturbative corrections. They are suppressed
by exponential factors of the form \eqn\expfin{ \exp(-e^{-K})=
\exp(-V_{T^6}/g_{top}^2) } Note that the exponent is proportional to
$1/g_{top}^2$ rather than $1/g_{top}$, which is reminiscent of the
instanton corrections in a gauge theory rather than D-instanton
effects in string theory.

The nonperturbative corrections in the OSV transform for the
$\CN=4$ black hole is discussed in Appendix A. We find that again
they behave as $\exp(-\alpha V/g_{top}^2)$, although unlike the
$\CN=8$ case, the exponent $\alpha$ is bounded both from above and
from below.

It would be interesting to understand the physical meaning of
these corrections, and in particular the possible connection to
the baby universe interpretation as in \DijkgraafBP. We should
note that the $\CN=8$ black hole degeneracy was derived in \ssyii\
while ignoring the effects due to fragmentation. Such effects
might need to be taken into account in order to accurately match
the nonperturbative corrections with multi-black hole states and
the corresponding interpretation in the topological string.

\vskip 0.8cm

\noindent {\bf Acknowledgments:}

We thank G.~Moore, A.~Neitzke, V.~Pestun, and A.~Strominger for
useful discussions. The research of DS is supported in part by an
NSF Graduate Research Fellowship and by NSF grant PHY-0243680. The
research of XY is supported in part by DOE grant DE-FG02-91ER40654.
Any opinions, findings, and conclusions or recommendations expressed
in this material are those of the author(s) and do not necessarily
reflect the views of the National Science Foundation.

\appendix{A}{Nonperturbative Corrections to the $\CN=4$ OSV
Transform}

In this appendix, we will analyze the corrections to the $\CN=4$
OSV transform, the $\dots$ in section 3. We will show that these
corrections are nonperturbative in $g_{top}$, justifying our
neglect of them in the text.

As discussed in section 3, these corrections come from the
residues of rational quadratic divisors other than \ratl.
Following the appendix of \DVV, the most general RQD is
characterized by five integers ${\bf v}=(k,l,m,a,c)$ satisfying
one constraint
\eqn\klmac{ 4ac-4kl+m^2-1=0 } The RQD is then
\eqn\klmacrqd{ a(\rho\sigma-\nu^2)+k\rho+l\sigma+m\nu+c=0 }
Substituting $\sigma={\phi^0\over2\pi i p^1}$, $\nu={\phi^1\over
2\pi i p^1}+{1\over2}$, we can solve for $\rho$:
 \eqn\rhosol{
\rho({\bf v}) = -{l\over a}+{((a-m)\pi p^1+i a \phi^1)^2-(\pi
p^1)^2 \over 2 a \pi p^1(2 k \pi p^1-i a\phi^0)}
}
The RQD \ratl\ corresponds to $a=m=1$, $k=l=c=0$, in which case
\rhosol\ reduces to the third equation of \defsss.

Repeating the calculation in section 3.1, we see that the general
RQD contributes an amount to $Z_{\CN=4}(p,\phi)$ which goes like
\eqn\subleadingrqd{
\delta Z_{\CN=4} \sim e^{\pi i \rho({\bf v})C_{MN}p^M p^N}
}
Let us now compare this with the contribution from \ratl, which
goes like $e^{\pi i \rho_* C_{MN}p^M p^N}$. We have
 \eqn\subleadingrqdcomp{
{\delta Z_{\CN=4}\over Z_{\CN=4}} = e^{\pi i(\rho({\bf
v})-\rho_*)C_{MN}p^M p^N}
 }
The real part of this exponent is
 \eqn\subleadingre{\eqalign{
\delta F &\equiv {\rm Re}\left[\pi i(\rho({\bf v})-\rho_*)
C_{MN}p^M p^N\right] \cr
 &= -{\pi ^2 C_{MN}p^M p^Np^1\over
2\phi^0}\left( { ((a-m)\phi^0+2k\phi^1)^2+(2\pi k
p^1)^2+(a^2-1)(\phi^0)^2\over (2\pi k p^1)^2+(a\phi^0)^2} \right)
 }}
This quantity is clearly negative, provided we assume that
${C_{MN}p^M p^N p^1\over\phi^0} > 0$ and take $a\ge 1$.

The condition $a\ge 1$ follows from the contour of $\rho$
integration. The contour must necessarily avoid the $a=0$ RQDs,
since otherwise the $a=0$ RQDs would dominate the exact
degeneracies and ruin the large charge asymptotics
$\Omega_{\CN=4}(p,q)\sim e^{S_{cl}(p,q)}$.

In fact, we are free to choose the contour such that only a single
$a=1$ RQD, say $k=l=c=0$ and $m=1$, contributes. The reason is
that the other $a=1$ RQDs are related to this one by integer
shifts of $\rho$, $\nu$, $\sigma$, and so they all contribute
identically to $Z_{\CN=4}(p,\phi)$. This is nontrivial from the
point of view of the above expressions; however, it is easy to see
if we write $Z_{\CN=4}(p,\phi)$ as
\eqn\czrew{ Z_{\CN=4} = {1\over
(p^1)^2}\sum_{k^0=0}^{p^1-1}\sum_{k^1=0}^{p^1-1}Z_{\CN=4}'(p,\phi^0+2\pi
i k^0,\phi^1+2\pi i k^1,\phi^M) } with \eqn\czpdef{
Z_{\CN=4}'(p,\phi)\equiv \sum_{q_M}
      \oint {d\rho\over\Phi(\rho,\sigma={\phi^0\over
      2\pi i p^1},\nu={\phi^1\over 2\pi i
      p^1}+{1\over2})}
       e^{\pi i \rho
       C_{MN}p^M p^N + {\phi^0\over 2p^1}C^{MN}q_M q_N+{\phi^1\over p^1}p^M q_M-\phi^M
       q_M}
 }
The $(a=1,k,l,m,c)$ RQD with $\sigma={\phi^0\over2\pi i p^1}$ and
$\nu={\phi^1\over2\pi i p^1}+{1\over2}$ can be brought to the form
$(a=1,0,0,1,0)$ by shifting $\phi^1$ and $\phi^0$ by $2\pi i p^1
s$ and $2\pi i p^1 t$ for some integers $s$ and $t$. Since
$Z_{\CN=4}'$ is clearly invariant under such shifts, this shows
that all the $a=1$ RQDs contribute identically to $Z_{\CN=4}$. By
choosing the contour so that only one $a=1$ RQD contributes, we
avoid overcounting.

Therefore, we have shown that aside from the contribution of the
$a=1$ RQD \ratl, the only other contributions to the black hole
partition function are from RQDs with $a>1$, and all of these are
nonperturbative. Let us now examine the $a>1$ RQDs in more detail.

Although we have considered the most general RQD, we do not expect
all of them to contribute to the exact degeneracies, for the
following reason. $\Phi$ is only holomorphic in the Siegel upper
half plane, defined by
\eqn\siegel{
 ||\Omega|| \equiv \det \,{\rm
Im}\,\left(\matrix{ \rho &\nu \cr \nu &\sigma } \right)
> 0
 }
Therefore, we should only consider the contributions of RQDs which
satisfy \siegel. Substituting once again $\sigma={\phi^0\over2\pi
i p^1}$, $\nu={\phi^1\over 2\pi i p^1}+{1\over2}$, and
$\rho=\rho({\bf v})$, we find a simple relation between $\delta F$
and $||\Omega||$:
\eqn\siegelrel{ \delta F = {2 \pi^2 C_{MN}p^M p^N p^1\over \phi^0}
\left(||\Omega||-{1\over4}\right)
 }
Therefore, the exponential suppression (a negative quantity) of
the $a>1$ RQDs is bounded from {\it below} by
\eqn\deltaFboundbelow{
\delta F>-{1\over4}{2 \pi^2 C_{MN}p^M p^N p^1\over \phi^0}
}
In addition, one can easily deduce from \subleadingre\ an {\it
upper} bound on $\delta F$ for $a>1$. Combining these, we conclude
that the exponential suppression of the subleading RQDs is
contained within a tight band:
\eqn\deltaFbound{
 -{\pi^2 C_{MN}p^M p^N p^1\over2\phi^0} < \delta F <
 -{3\over 4}\times {\pi^2 C_{MN}p^M p^N p^1\over2\phi^0}
}
The lower bound on $\delta F$ is intriguing. It should be
contrasted with the very different behavior of the nonperturbative
corrections in the $\CN=8$ case \intfuls, where we found a series
of successively smaller nonperturbative corrections.

 \listrefs
\end